\documentclass{ieeetj}
\usepackage{cite}
\usepackage{amsmath,amssymb,amsfonts}
\usepackage{algorithmic}
\usepackage{hyperref}
\hypersetup{hidelinks=true}
\usepackage{algorithm,algorithmic}
\usepackage{graphicx}
\usepackage{textcomp}
\usepackage{wrapfig,colortbl}
\usepackage{dblfloatfix}
\usepackage{tabularx}
\usepackage[table]{xcolor}
\let\labelindent\relax
\usepackage{enumitem}

\newcommand{\figurePath}{.}

\def\BibTeX{{\rm B\kern-.05em{\sc i\kern-.025em b}\kern-.08em
    T\kern-.1667em\lower.7ex\hbox{E}\kern-.125emX}}
\AtBeginDocument{\definecolor{tmlcncolor}{cmyk}{0.93,0.59,0.15,0.02}\definecolor{NavyBlue}{RGB}{0,86,125}}

\def\OJlogo{\vspace{-4pt}$<$Society logo(s) and publication title will appear here.$>$}
\def\seclogo{\vspace{10pt}$<$Society logo(s) and publication title will appear here.$>$}

\def\authorrefmark#1{\ensuremath{^{\textbf{#1}}}}

\begin{document}

\receiveddate{XX Month, XXXX}
\reviseddate{XX Month, XXXX}
\accepteddate{XX Month, XXXX}
\publisheddate{XX Month, XXXX}
\currentdate{XX Month, XXXX}
\doiinfo{XXXX.2025.1234567}

\markboth{}{Dahunsi {et al.}}

\title{Hadamard Encoded Row Column Ultrasonic Expansive Scanning (HERCULES) with Bias-Switchable Row-Column Arrays }
\author{Darren Dahunsi\authorrefmark{1}, \IEEEmembership{Student Member, IEEE}, 
Randy Palamar\authorrefmark{1}, \IEEEmembership{Student Member, IEEE},
Tyler Henry\authorrefmark{1}, \IEEEmembership{Student Member, IEEE}, 
Mohammad Rahim Sobhani\authorrefmark{1,3}, \IEEEmembership{Student Member, IEEE}, 
Negar Majidi\authorrefmark{1}, \IEEEmembership{Student Member, IEEE}, 
Joy Wang\authorrefmark{1}, \IEEEmembership{Student Member, IEEE}, 
Afshin Kashani Ilkhechi\authorrefmark{1,3}, \IEEEmembership{Student Member, IEEE}, 
Jeremy Brown\authorrefmark{2}, \IEEEmembership{Member, IEEE}
 and 
Roger Zemp\authorrefmark{1,3}, \IEEEmembership{Member, IEEE}
}
\affil{Department of Electrical and Computer Engineering, University of Alberta, Edmonton, AB T6G 2R3, Canada}
\affil{School of Biomedical Engineering, Dalhousie University, Halifax, NS B3H 4R2, Canada}
\affil{CliniSonix Inc., Edmonton, AB  T5J 4P6, Canada}
\corresp{Corresponding author: Darren Dahunsi (email: dahunsi@ualberta.ca).}
\authornote{RJZ and MRS are directors and shareholders of CliniSonix Inc., which provided partial support for this work. RJZ is a founder 
and director of OptoBiomeDx Inc., which, however, did not support this work. RJZ is also a founder and shareholder of IllumiSonics Inc.,which, however, did 
not support this work. JB is a shareholder and director of SoundBlade Inc.,and DaxSonics Inc. which, however, did not support this work.
    We gratefully acknowledge funding from the National Institutes of Health  (1R21HL161626-01 and EITTSCA R21EYO33078), Alberta Innovates (AICE 202102269, 
CASBE 212200391 and LevMax 232403439), MITACS (IT46044 and IT41795), CliniSonix Inc., NSERC (2025-05274) and a NSERC CGS-D to DD (588860 - 2024), the Alberta Cancer Foundation and the Mary 
Johnston Family Melanoma Grant (ACF JFMRP 27587), the Government of Alberta Cancer Research for Screening and Prevention Fund (CRSPPF 017061), an Innovation 
Catalyst Grant to MRS, INOVAIT (2023-6359), and an IBET Momentum Fellowship to DD. We are grateful to the nanoFAB staff at the University of Alberta for 
facilitating array fabrication.}

\begin{abstract}
    Top-Orthogonal-to-Bottom-Electrode (TOBE) arrays, also known as bias-switchable row-column arrays (RCAs), allow for imaging techniques otherwise impossible
    for non-bias-switachable RCAs. Hadamard Encoded Row Column Ultrasonic Expansive Scanning (HERCULES) is a novel imaging technique that allows for 
    expansive 3D scanning by transmitting plane or cylindrical wavefronts and receiving using Hadamard-Encoded-Read-Out (HERO) to perform beamforming 
    on what is effectively a full 2D synthetic receive aperture. This allows imaging beyond the shadow of the aperture of the RCA array, potentially 
    allows for whole organ imaging and 3D visualization of tissue morphology. It additionally enables view large volumes through limited windows. In
    this work we demonstrated with simulation that we are able to image at comparable resolution to existing RCA imaging methods at tens to hundreds of 
    volumes per second. We validated these simulations by demonstrating an experimental implementation of HERCULES using a custom fabricated TOBE 
    array, custom biasing electronics, and a research ultrasound system. Furthermore, we assess our imaging capabilities by imaging a commercial 
    phantom, and comparing our results to those taken with traditional RCA imaging methods. Finally, we verified our ability to image real tissue by 
    imaging a xenograft mouse model.
\end{abstract}


\maketitle

\section{Introduction}
\label{sec:introduction}\IEEEPARstart{U}{ltrasound}, as a low-cost, non-invasive, accessible imaging modality  can enable various diagnostic and clinical applications. 
Despite this flexibility, ultrasound usability is limited by the difficulty in properly assessing and estimating 
features while relying only on a 2-dimensional view. This creates a reliance on trained operators, as wells as operator-dependent variance in measurements.

\begin{figure*}
    \centerline{\includegraphics[width=\linewidth]{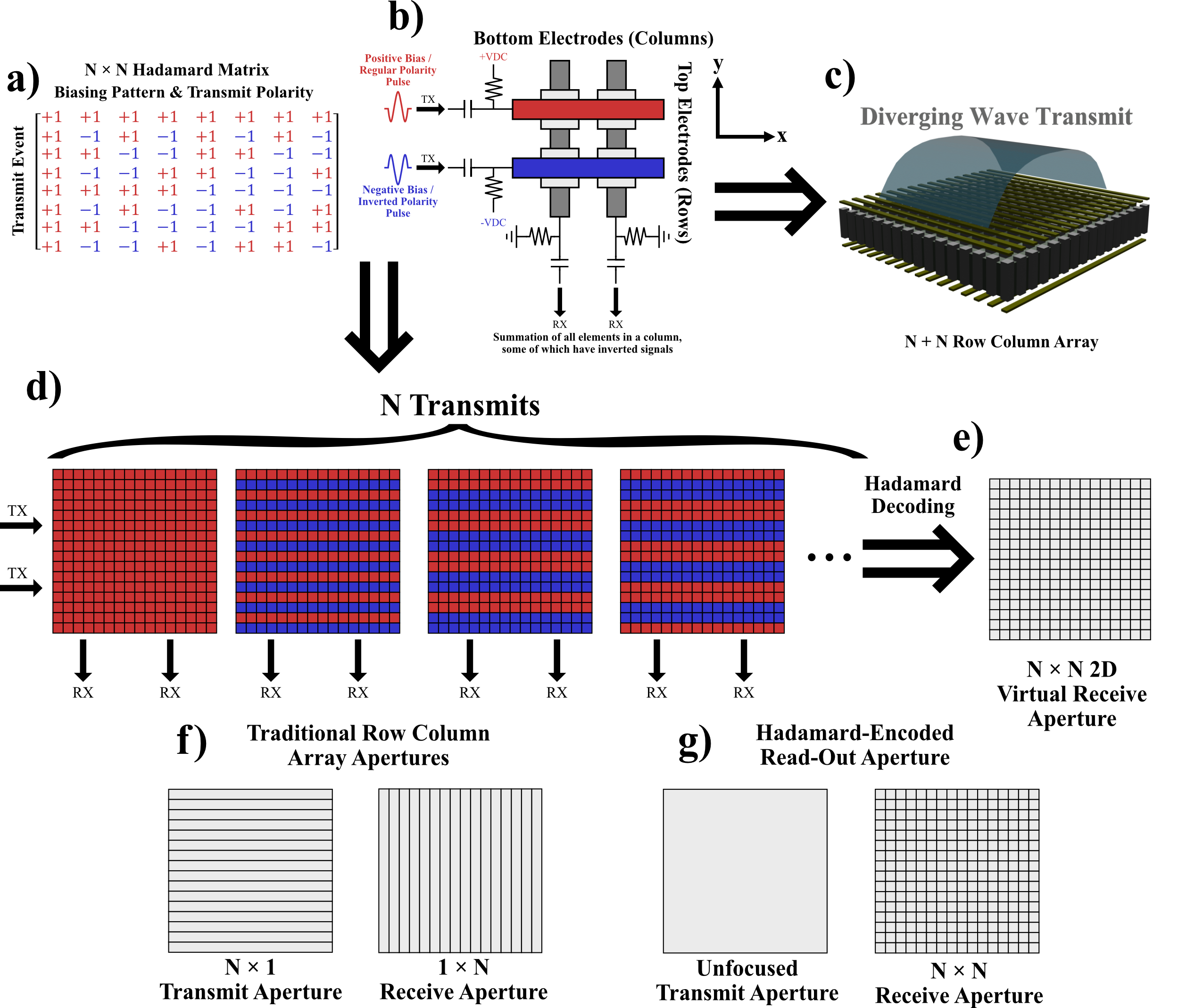}}
    \caption{Summary of the HERCULES imaging scheme for an arbitrarily sized array of N rows and N columns. In a) we use the rows of a hadamard matrix to 
    determine the biasing and transmit polarities that are applied on the rows as shown in b), with each column corresponding to a different row. 
    c) shows a render of the effective layout of the element electrodes on the transducer, and the effective transmit waveform. In d) we can see the effective 
    bias pattern between each transmit event. While the transmitted waveform will not change during transmits, the received waveform will, and decoding will 
    result in the receive aperture shown in  e). f) \& g) show the differences between the effective apertures of a traditional Row-Column Array imaging method
    and the HERCULES imaging scheme.}
    \label{fig:imagingScheme}
\end{figure*}

These problems have prompted interest in 3D ultrasound imaging \cite{Smith:1991:HighSpeedSystemA, Welch:2001:Freehand3D, Austeng:2002:Sparse2DArray3DImaging,
Yen:2004:3DRectilinearMultiplex, Chang:2015:GraphSegmentation3D}. Volumetric imaging can also better enable other ultrasound applications such as flow imaging, 
elastography, and super-resolution imaging \cite{Downey:1995:3DVascularDoppler, Huang:2015:3DLinearStrain}.  A wide spectrum of information in a 3D field of 
view could give clinicians a comprehensive overview of organ health and operation, with a non-invasive, and accessible imaging modality.

To achieve this goal, better visualize tissue morphology and perform full organ imaging with ultrasound, there has been increased research in the development of 
3D ultrasound imaging. Various methods have been used, but most methods are subject to various limitations. Mechanical scanning makes use of traditional 
ultrasound arrays with their proven imaging characteristics in a 2D plane, but this tends to create motion artifacts, and framerate is limited due to physical 
constraints of mechanical systems \cite{Lockwood:1998:3DSparseSA}. Fully wired matrix probes can image with high resolution and contrast, but require complex 
wiring and high element counts and are thus limited in size and resolution. Additionally, they often need to make use of microbeamformers which introduce 
imaging artifacts \cite{Smith:1991:HighSpeedSystemA, vonRamm:1991:HighSpeedSystemB}.

Row-Column addressed Arrays (RCAs) are able to image in a wide field of view at high framerates, but are unable to simultaneously focus in both the azimuthal 
and elevational directions \cite{Seo:2009:RCA, Rasmussen:2013:3DRCASimulation, Wong:2014:MicromachinedRCA}. This creates asymmetrical resolution properties that
can be worse than matrix probes, but they require far lower element counts.  As opposed to matrix arrays whose element count scale quadratically with the size 
of the array, RCAs scale linearly, as they are effectively two linear arrays stacked orthogonally on top of each other. Various acquisition strategies have been
used to mitigate the disadvantages of RCAs, most notably coherently compounding multiple acquisitions with different virtual sources 
\cite{Awad:2009:3DSpatialCompoundRCA, Flesch:2017:4DUltrafastRCA,Jørgensen:2023:RCABeamformer}.

Volumetric scanning with RCAs creates additional challenges.  RCA imaging schemes typically try to take advantage of the benefits presented by ultrafast 
ultrasound techniques \cite{Shattuck:1984:Explososcan, Jain-Yu:1997:LimitedDiffractionBeams, Jain-Yu:1998:LimitedDiffractionBeamsExperiments, 
Flesch:2017:4DUltrafastRCA, Sauvage:2018:4DUltrafastDopplerRCA}. One of the most significant drawbacks of ultrafast ultrasound imaging with linear arrays is the 
lack of elevational focusing, but RCAs can mitigate this issue by transmitting with the rows of the arrays while receiving ultrasound signals with the columns, 
as show in Figure \ref{fig:imagingScheme} (f) \cite{Jensen:2022:RCALinearComparison}.  However, this creates a new problem; while one aperture can project 
energy in a wide field in a plane, the other will lack any focusing ability in that direction and thus limit the field of view to below the shadow of the 
aperture. This limits the feasibility of RCAs for clinical applications requiring limited viewing windows but large volume requirements such as cardiac and 
intracranial imaging \cite{Jensen:2022:AnatomicRCA, Wu:2024:3DTranscranialDynamicUltrasound}. Diverging acoustic lenses can be used to obtain a wider field
of view, but RCAs would still be limited to a cylindrical swath.

Electrostrictive transducers only act like piezoelectrics when a DC bias voltage is applied across them. This allows for parts of the array to be selectively 
'turned on and off'. This property has unique applications when applied to row-column arrays, where the two linear arrays are interconnected; choosing to 
selectively bias the rows can enable or disable portions of each column and vice versa. Bias-sensitivity opens up the possibility of encoding additional 
information into transmit sequences, also known as bias-encoding.  At an extreme every row-column element pair combination can potentially be addressed to 
create a unique aperture. Electrostrictive materials also demonstrate that the polarity of transmitted and received ultrasound signals invert depending on the 
polarity of the DC bias across the electrodes \cite{Sherrit:1998:Electrostrictive}. Combined with the general linearity of small scale ultrasound signals, this
can be leveraged to achieve higher average signal-to-noise-ratios while performing bias encoding \cite{Ceroici:2017:Forces, Ceroici:2019:ForcesExp, 
Sobhani:2022:uForces}.

\begin{figure}
    \centerline{\includegraphics[width=\columnwidth]{\figurePath/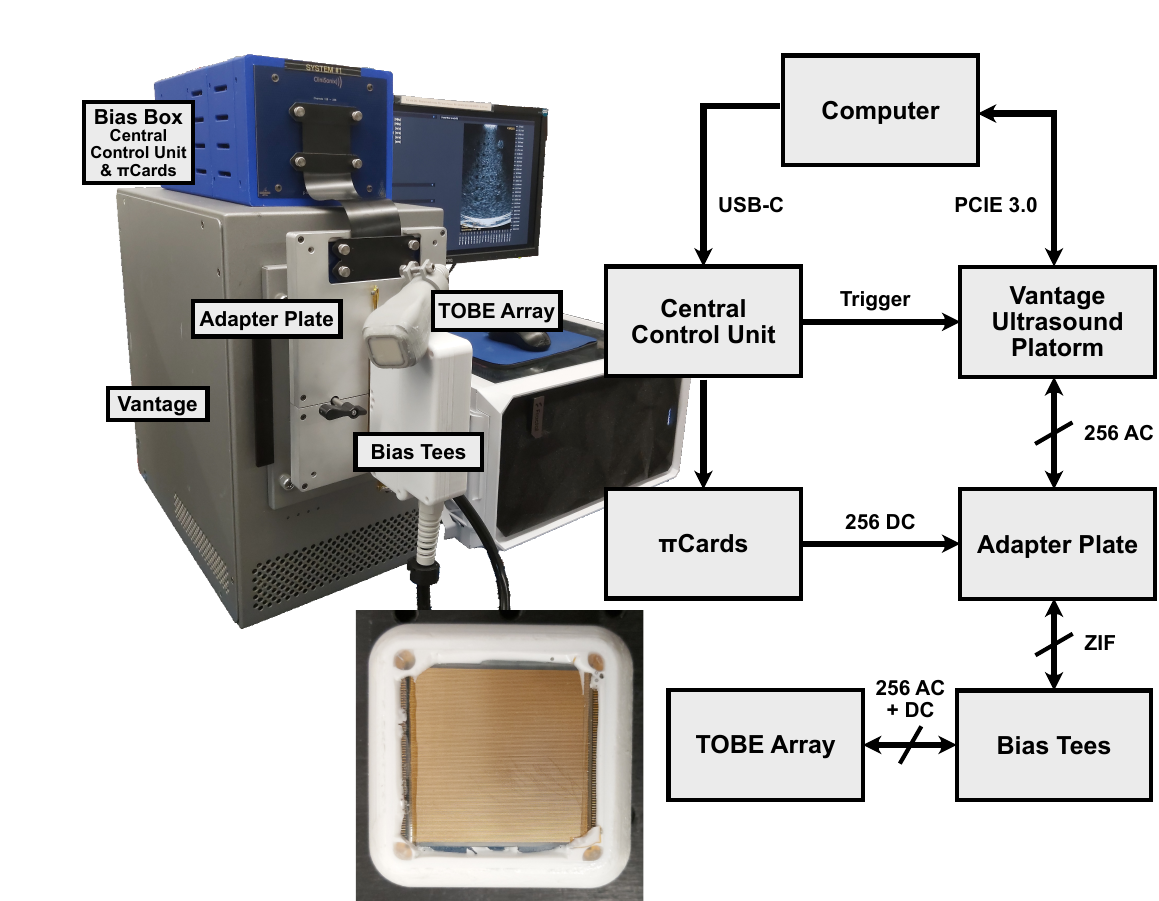}}
    \caption{Experimental setup for phantom and in-vivo imaging. A computer controls both a Vantage Ultrasound Research platform, and a central control unit for 
    custom biasing electronics. The Central Control Unit controls the biasing cards ($\pi$Cards), as well as sends a trigger signal to Vantage system to notify
    it of biasing events.  The Vantage systems' AC signals are mixed with the $\pi$Cards DC signals on a bias tee as sent to the transducer. A picture of an
    unhoused transducer is also shown.}
    \label{fig:experimentalSetup}
\end{figure}

In a previous work we demonstrated that we can use electrostrictive bias encoding to achieve transmit encoding and create an effective synthetic transmit 
aperture. This effective aperture allows us to make high resolution 2D images, that have the potential to even surpass the imaging quality of matrix probes in a 
2D plane \cite{Ceroici:2017:Forces}. We have also shown in previous work that we can perform volumetric photoacoustic imaging with RCAs by using selective 
biasing \cite{Ceroici:2018:Photoacoustic}. In this work we will build on these achievements by demonstrating that we can perform volumetric ultrasound imaging 
using bias-sensitive row-column arrays. Using bias encoding, we will create a virtual synthetic receive aperture reminiscent of a fully wired matrix probe. 
Furthermore, unlike traditional RCA imaging methods, this imaging scheme will be able to use cylindrically diverging waves to imaging beyond the shadow of the 
aperture of the probe.  This imaging scheme is call Hadamard Encoded Row Column Ultrasonic Electronic Scanning (HERCULES).

\section{Methods}

\subsection{Hadamard-Encoded Read-Out (HERO) Acquisition}

For the follow description we will describe the rows as spanning the x-axis and varying in position along the y-axis, while the columns span the y-axis and vary
in position along the x-axis as shown in Figure \ref{fig:imagingScheme} b). Additionally, the x-axis corresponds to the lateral direction, and the y-axis 
corresponds to the elevational direction.

Let us model a typical row-column array by subdividing the array into a grid of 'physical elements' delineated by the overlap of row and column electrodes. 
Assuming the transmit pattern is constant for every transmit event $e$, we will call the signal at row $r$ and column $c$ $s_{rc}(t)$. 

If we receive from columns, while biasing the rows with a bias pattern selected from a Hadamard matrix $H$, the column signals for a transmit event $e$ can be 
modeled as:

\begin{equation}
    g_c^{\{e\}}(t) = \sum_{r} H_r^{\{e\}}s_{rc}(t)  
\end{equation}

\begin{table}[t]
    \caption{Parameter Used in Field II Simulations}
    \label{table:simDieParams}
    \setlength{\tabcolsep}{3pt}
    \begin{tabular}{p{150pt}p{75pt}}
        \hline\hline
        Parameter                       &
        Value                           \\
        \hline
        Speed of Sound                  &
        1540 m/s                        \\
        Center frequency                &
        6.3 MHz                           \\
        Sampling frequency              &
        50 MHz                          \\
        Pitch                           &
        250 $\mu$m                      \\
        Kerf                            &
        30 $\mu$m                       \\
        Number of excitation cycles     &
        1                               \\
        2D array size                   &
        128 $+$ 128                     \\
        Transmit Count                  &
        128                             \\
        Point Target Depth              &
        50 mm                           \\
        \hline\hline
    \end{tabular}
\end{table}

This may be written in matrix form as $\mathbf{g = Hs}$. After acquiring from the complete set of transmit events, the channel data from every element can then 
be estimated as $\hat{s} = H^{-1}g$. This procedure constitutes the Hadamard-Encoded Read-Out (HERO) acquisition scheme. This was previously used for 
photoacoustic readout, but not yet for volumetric ultrasound imaging \cite{Ceroici:2018:Photoacoustic}. Note that the row biasing patterns may alternate 
polarity, which for electrostrictive materials will result in an inversion of the transmit waveform. To ensure that the same transmit pattern is emitted for 
every transmit event, we must account for this inversion. Our approach thus inverts the polarity of the transmit waveform for rows where the biases are negative 
\cite{Ceroici:Thesis}.

As an example, for an 128 + 128 channel TOBE array, we can use 128 transmit events to obtain the effective results of a diverging wave transmit and full 
synthetic receive acquisition from a fully wired 128 $\times$ 128 matrix array. Figure \ref{fig:imagingScheme} illustrates an arbitrary N-transmit HERO 
acquisition. If we use a virtual source behind the array we can generate a diverging cylindrical wave, and as our receive aperture is a 2D array, we can focus 
everywhere in receive. We can repeat this after swapping the roles of the rows and columns, and image a notched pyramidal volume.

\begin{figure*}[t]
    \centerline{\includegraphics[width=0.8\linewidth]{\figurePath/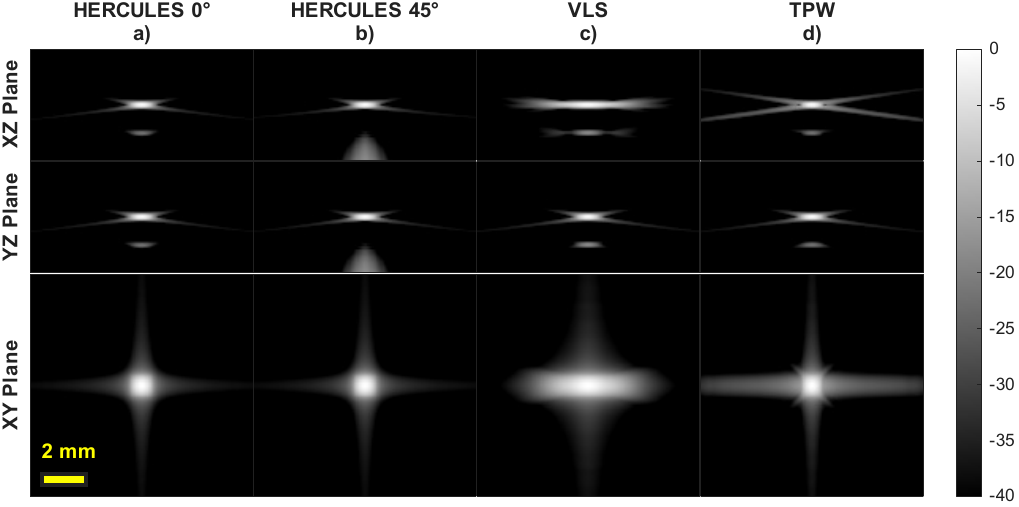}}
    \caption{Simulation of point target located 50 mm below the center for the array.  For each imaging method we show the Point-Spread-Functions (PSFs) in the 
    lateral-axial, elevational-axial, and lateral-elevational planes, centered on the point target. Below are lateral, elevational, and axial profile 
    measurements.}
    \label{fig:simulatedPsf}
\end{figure*}

\subsection{Simulation}

\begin{figure}
    \centerline{\includegraphics[width=0.8\columnwidth]{\figurePath/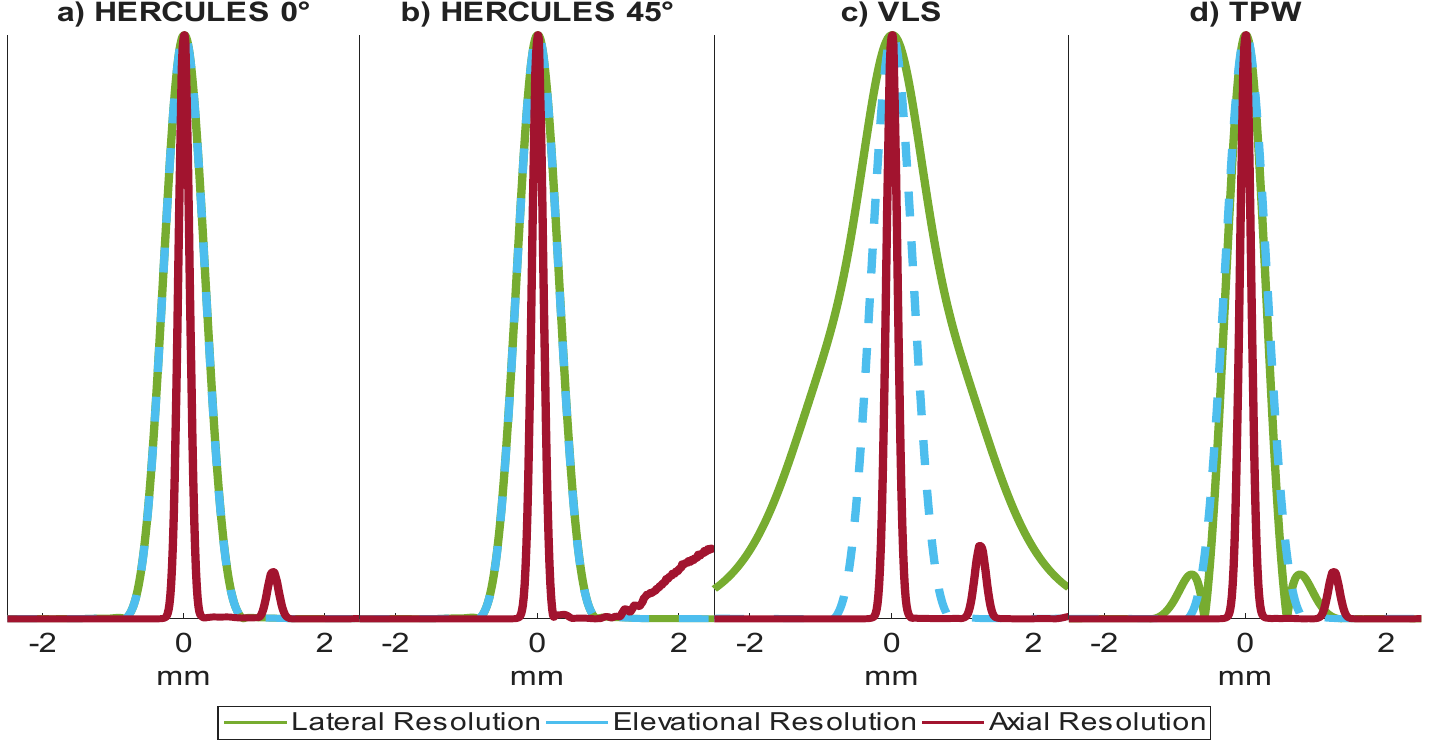}}
    \label{fig:simulatedPsfProfile}
\end{figure}

We hypothesize that HERCULES will provide imaging characteristics that are comparable to traditional row-column array, with the potential to also result in a
reduction in artifacts.  Additionally, we expect that we can achieve comparable SNR when imaging a point outside the shadow of the aperture as when imaging a 
point under the shadow if we use a diverging wave. To this end, we will compare 3 imaging schemes along with some variations in \textit{silico}. These imaging 
schemes are: 1) RCA imaging using a walking virtual line source (VLS) for transmission \cite{Rasmussen:2015:3DRCA1}. 2) RCA imaging using tilted plane 
wave (TPW) transmissions \cite{Chen:2018:RCA3DPW}. 3) HERCULES imaging.

We use Field II \cite{FieldII:Theory,FieldII:Program} to simulate point spread functions for each imaging scheme. We model our RCAs as a 2D array of physical 
elements and apply an apodization reflecting the bias states of the intersecting row \& column. Receive signals are created by adding up the signals of
every element for a given row or column. For a fair comparison we utilize the same array specifications for all simulations; these parameters are listed in 
Table~\ref{table:simDieParams}. Additionally, we maintain a constant number of transmits for a consistent acquisition time and energy output.

\begin{table}
    \caption{Simulation Measurements}
    \label{table:simResults}
    \setlength{\tabcolsep}{3pt}
    \begin{tabular}{p{56pt}p{40pt}p{40pt}p{40pt}p{40pt}}
        \hline\hline
        Acquisition                     &
        HERCULES Planewave (0°)         &
        HERCULES Diverging (45°)        &
        Virtual Line Source             &
        Tilted Planewave                \\
        \hline
        Lateral Resolution ($\mu$m)     &
        675                             &
        675                             &
        1713                            &
        625                             \\
        Elevational Resolution ($\mu$m) &
        675                             &
        675                             &
        675                             &
        675                             \\
        Axial Resolution ($\mu$m)       &
        175                             &
        187                             &
        175                             &
        175                             \\
        CNR                             &
        0.6449                          &
        0.4094                          &
        0.2484                          &
        0.3436                          \\
        \hline\hline
    \end{tabular}
\end{table}

For this comparison we evaluate various imaging parameters. We measure the lateral, elevational, \& axial resolutions via the respective full-width-half-maximum 
(FWHM) value. To obtain an effective measure of the relative contrast to noise ratio (CNR) between the different imaging schemes, we will measure the ratio of 
energy within 2.5 wavelengths of the center of the PSF to the energy outside that radius. These measurements will later be further corroborated 
with contrast measurements on experimental data. (Figure \ref{fig:simulatedPsf}) shows simulated Point-Spread-Functions (PSFs) for the aforementioned imaging 
schemes.

\begin{figure*}
    \centerline{\includegraphics[width=\linewidth]{\figurePath/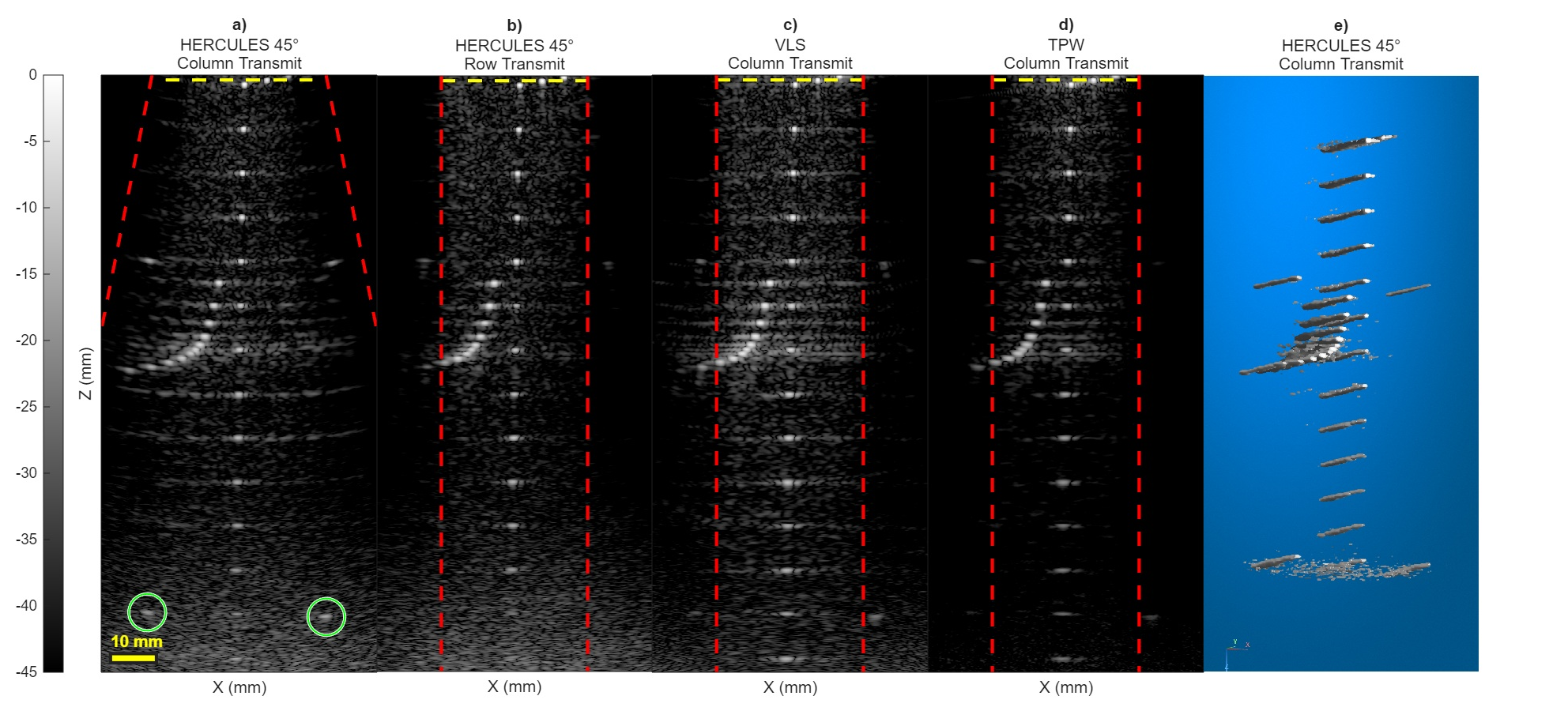}}
    \caption{Intensity image of resolution targets in a CIRS ATS 539 tissue mimicking phantom. Elevational slice of reconstructed 3D volume. In a) we can see 
    that our field of view is expanded. We can see in a) that at large depths some points outside the shadow of the aperture that would normally be barely 
    visible are clearly seen with HERCULES 45°. e) shows a volumetric render of the resolution targets.}
    \label{fig:resolution}    
    \centerline{\includegraphics[width=\linewidth]{\figurePath/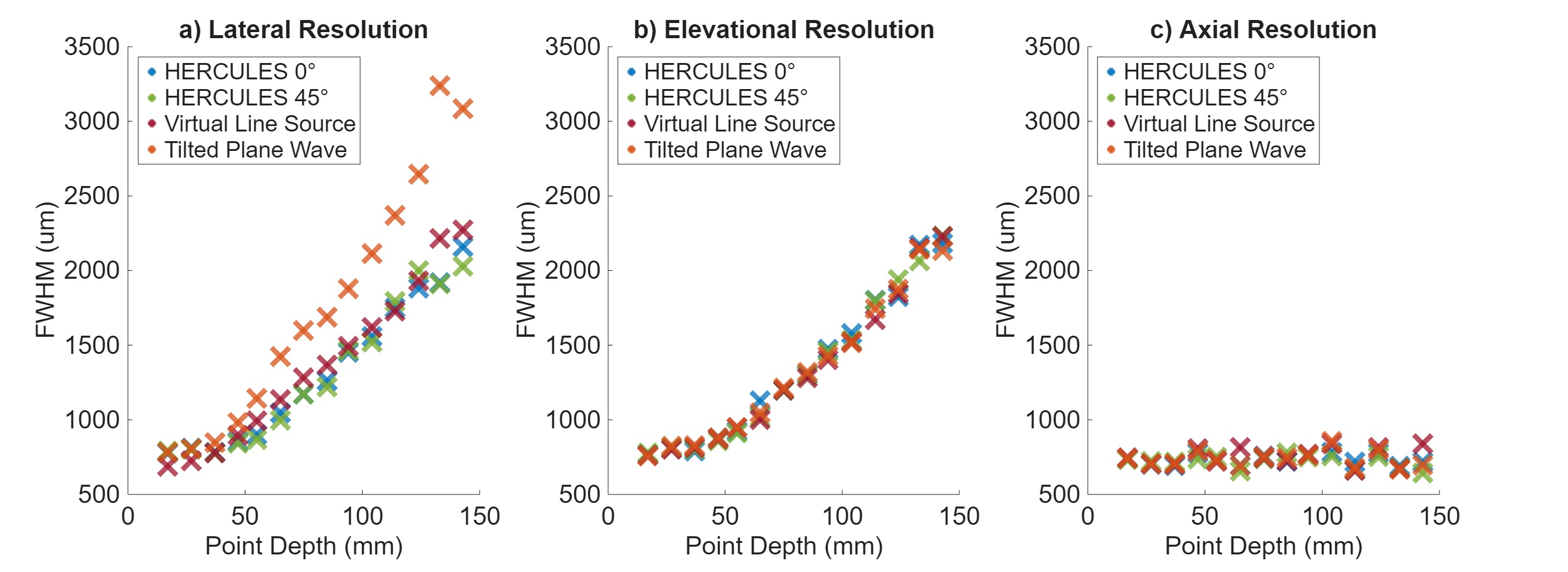}}
    \caption{Experimental measures of the resolution. Lateral resolution (a)) measurements are by transmitting on the columns and measuring the resolution in 
    the x-axis, while elevational resolution (b)) is measured by transmitting on the rows. Axial resolution (c)) is an average of the resolution in the z-axis 
    of the two acquisitions.}
    \label{fig:resolutionMetric}

\end{figure*}

\subsection{Experiment}

To experimentally evaluate and compare these imaging schemes we use custom fabricated TOBE arrays (CliniSonix, Edmonton, AB, Canada) 
\cite{Sampaleanu:2014:CMUT_TOBE_Ultrasound, Chee:2014:CMUT_TOBE_Photoacoustic}. These arrays can act as standard row column arrays, when one side is uniformly 
biased, and will thus allow for a consistent and fair comparison between standard RCA imaging methods and our bias-encoded imaging methods. Our TOBE arrays 
require that bias voltages be provided to individual channels independently.  We will use custom biasing electronics to do this 
\cite{Ilkhechi:2023:BiasSwitching}. Ultrasound transmit signals will be provided by an attached Vantage 256 High Frequency Ultrasound Research System 
(Verasonics, Kirkland, WA, USA). This system will also be responsible for receiving ultrasound signals. DC bias signals from the biasing electronics, and AC 
ultrasound signals from the Vantage System will be coupled together on bias tees and sent to the array. This is shown in Figure \ref{fig:experimentalSetup}.
To demonstrate the structural imaging capabilities of these imaging methods we will use low-to-mid frequency arrays, suitable for large organ imaging. 
Specifically, we will use a low-frequency 3-6 MHz 45$\times$45 mm and a mid-frequency 4-8 MHz 32$\times$32 mm TOBE handheld probe. To calculate imaging parameters we will image an ATS 539 
commercial quality assurance phantom (CIRS Inc., Norfolk, VA, USA). We seek to evaluate the resolution, CNR, and viewing angles of these different imaging 
methods, as well as their consistency across the imaged volume.

\begin{figure*}
    \centerline{\includegraphics[width=\linewidth]{\figurePath/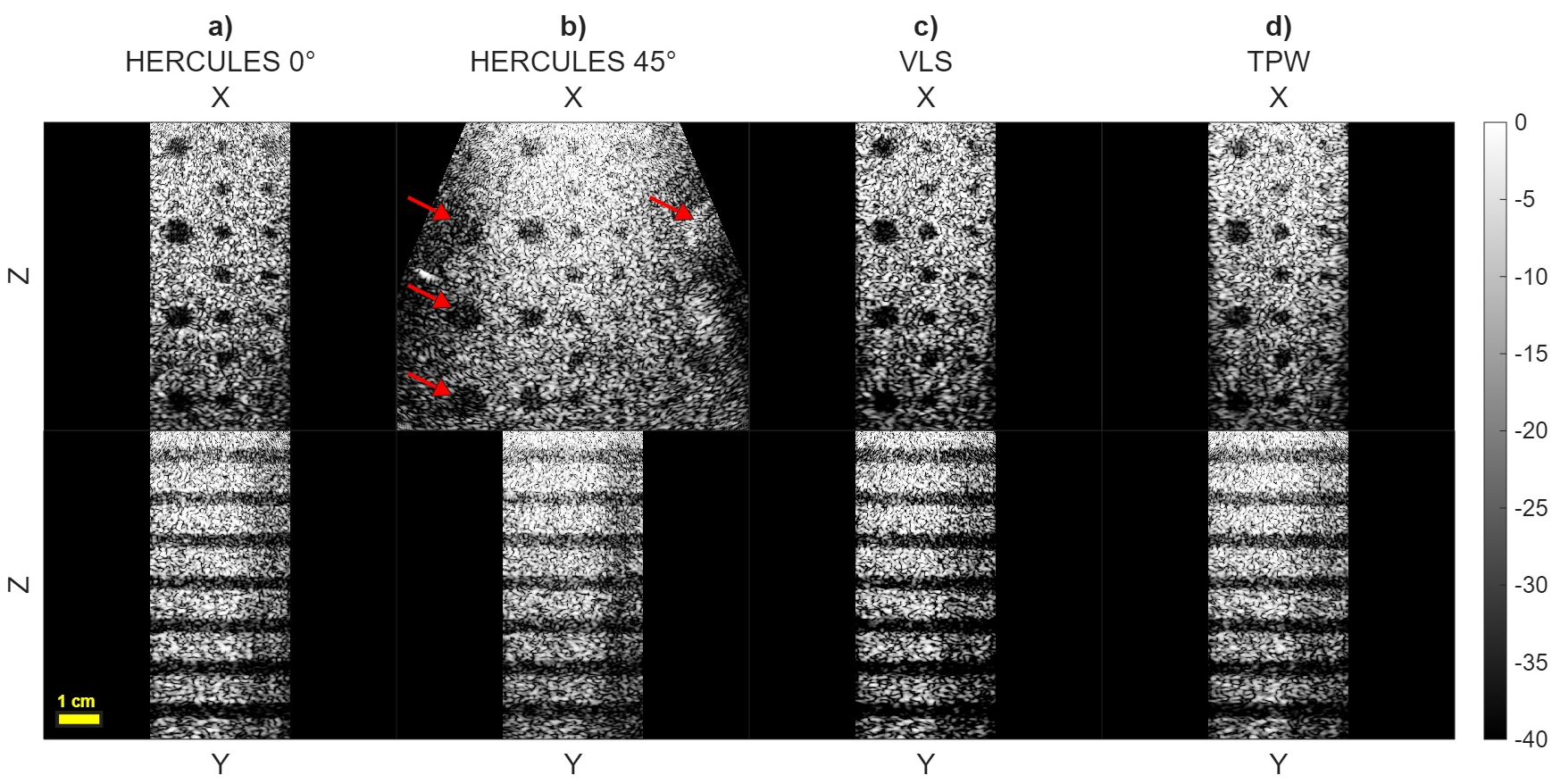}}
    \caption{Intensity image of cyst targets in a CIRS ATS 539 tissue mimicking phantom. 30 $\mu$s chirp transmits were used to increase penetration. 
    Elevational slice of reconstructed 3D volume. Each of the images is normalized relative to its maximum intensity. Hypoechoic cysts outside the shadow of 
    the aperture are only visible in b), where we utilize HERCULES with a diverging cylindrical wave.}
    \label{fig:cyst}
\end{figure*}

Compared to conventional focused imaging, ultrafast imaging suffers from a natural disadvantage; any given part of the imaged medium, is insonified to a far 
lesser degree, and thus the resulting image tends to suffer from lower SNR \& CNR \cite{Jensen:2006:SyntheticAperture,Nikolov:2006:SyntheticApertureChallenges}. 
To counteract this, we can lengthen the transmitted pulse, but this will potentially reduce our resolution. When imaging low contrast targets, we will use chirp 
pulses as these provide the long pulses we need, but can be filtered to nearly fully recover the lost resolution \cite{Misaridis:2005:CodedExcitationA,
Misaridis:2005:CodedExcitationB, Misaridis:2005:CodedExcitationC}. For volumetric imaging of the cyst targets we used 30$\mu$s long chirps with a bandwidth 
matching the bandwidth of the transducer.

To further evaluate the different imaging schemes, we performed imaging ex-vivo in mice. Tumor-bearing mice were used for these experiments. Animal experiments 
were approved through the University of Alberta ACUC (AUP \#3982 and \#2994). SCID Hairless Outbred (SHO, Charles River) mice were injected with subcutaneous 
B16F10 mouse melanoma tumors in the hind flank. Once flank tumors had grown to a size of at least 5 mm in diameter, the hind section of the mice were then 
submerged in degassed water and imaged with our 8MHz TOBE arrays.

\begin{figure}
    \centerline{\includegraphics[width=\columnwidth]{\figurePath/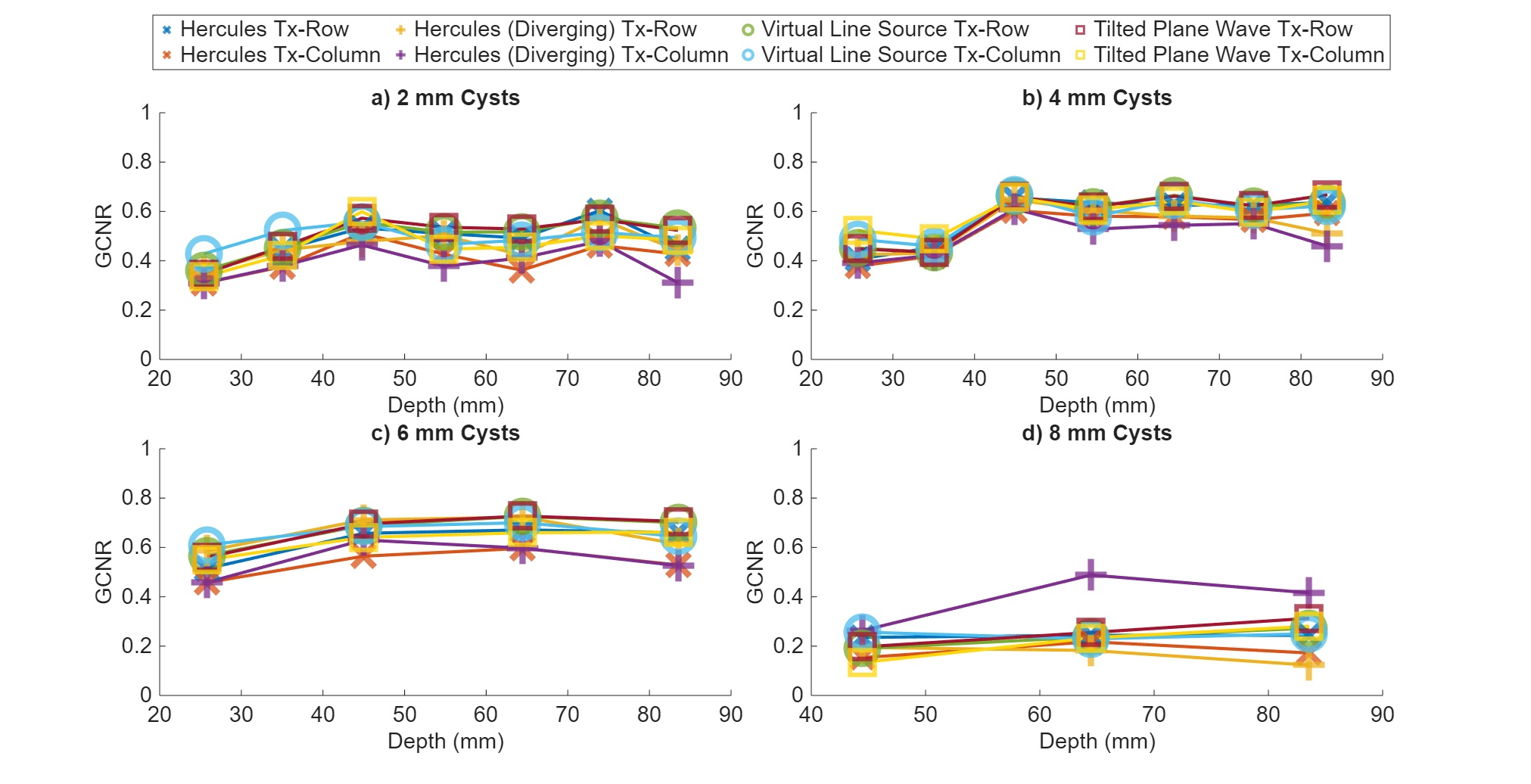}}
    \caption{ GCNR measurements of cysts shown in Figure \ref{fig:cyst}. Subfigures a-d) refers to each vertical column of cysts from the right to left, smallest
    to largest, respectively.}
    \label{fig:cystMetrics}
\end{figure}

\subsection{Processing}

We use intensity modulated ultrasound receive data from the Vantage System, and manipulate it with the rest of our beamforming 
pipeline to reconstruct our images. The first step is to apply a matched filter that models our transmit waveform to allow us to recover the effective 
results of a single cycles transmit. We also apply the Hilbert transform to obtain the analytical signal, and improve the effects of compounding in later 
processing stages. For our HERCULES data we then multiply the inverse of our encoding Hadamard matrix across transmits, to recover acquisition data for every 
element of our 2D grid of receive elements. From this point we perform standard delay-and-sum beamforming\cite{Fink:1992:1:TimeReversalA} to reconstruct an 
intensity image. We generate low resolution images from every transmit and every receive element, and coherently compound these images together to obtain a 
final high resolution image.
During experiments, we perform this beamforming at real-time imaging rates. We perform offline volume reconstruction with custom beamforming software and scripts 
to validate, and perform post-processing on our results. 2D images are normalized to their maximum values. 3D Volumes are rendered in MATLAB \cite{MATLAB}, 
with smoothing and thresholding.

To measure resolution we use the line targets present in the ATS 539 phantom. We cut an elevational slice from the imaged volume, and measure the lateral and 
axial resolution of the points in the image. Similarly, to measure contrast we use the anechoic cyst targets in the phantom. For our contrast metric we use
generalized Contrast-to-Noise-Ratio (gCNR) \cite{Rodriguez-Molares:2020:gCNR} and compare histograms of the pixels inside and outside the cysts to measure their 
detectability. In this case we beamform rectangular volumes centred on the cyst, elongated in the y-direction, and extract concentric cylindrical regions  
that represent inside, and outside the cyst, without including the transition region.

\begin{figure*}
    \centerline{\includegraphics[width=\linewidth]{\figurePath/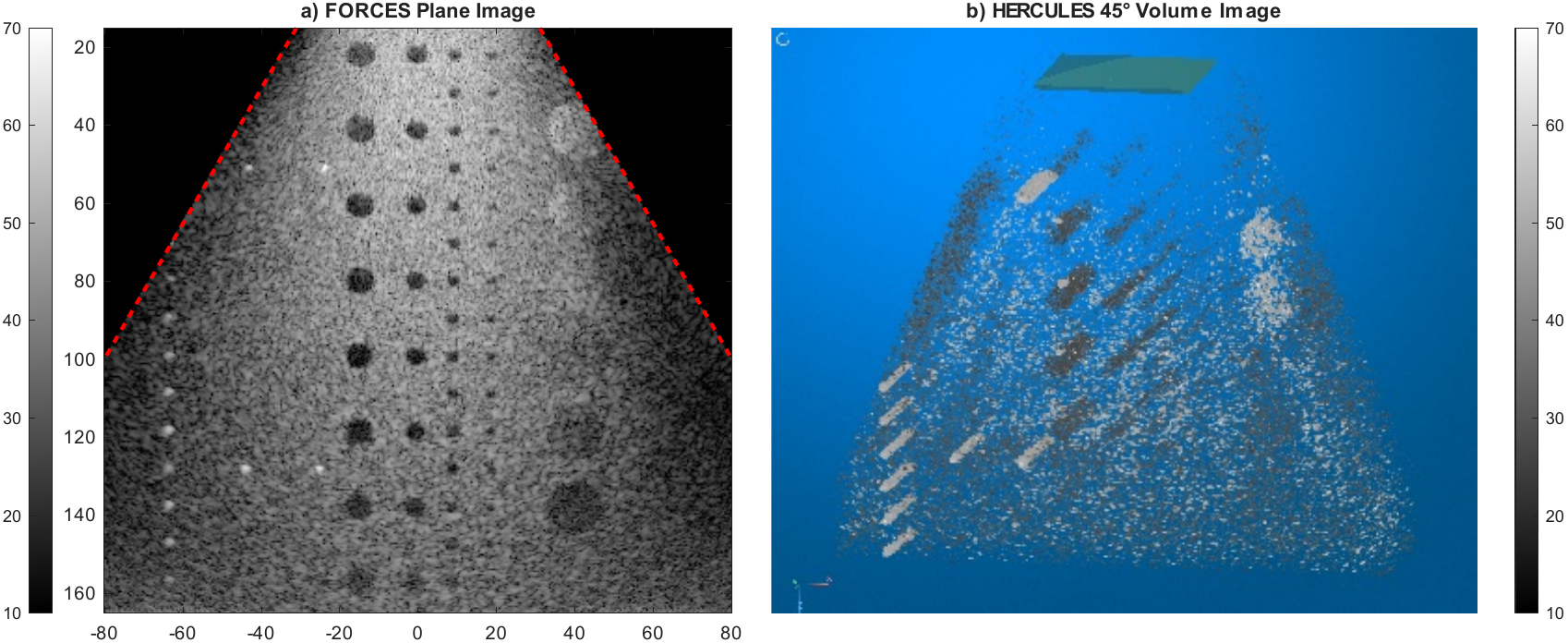}}
    \caption{Wide Field of View Images of CIRS ATS 539 tissue mimicking phantom using a low frequency TOBE array. In a), 
    Fast-Orthogonal-Row-Column-Electronic-Scanning (FORCES) is used to capture a high quality B-Mode image, while in b) HERCULES is used to acquire a lower 
    quality volume in a similar region. Refer to the supplementary material for a video showing cross-section slices of the volume.}
    \label{fig:phantomVolume}
\end{figure*}

\begin{figure*}[t]
    \centerline{\includegraphics[width=\linewidth]{\figurePath/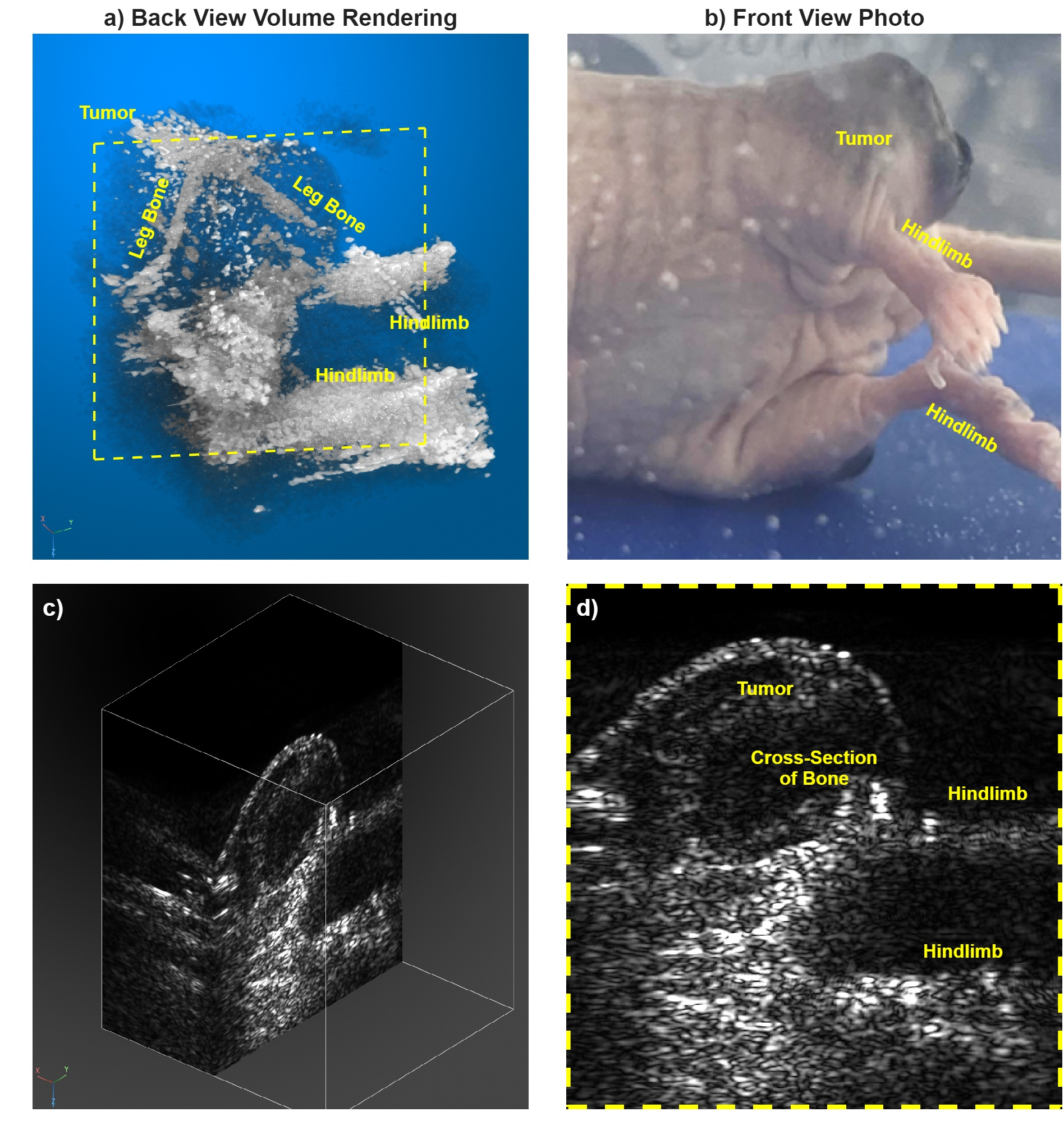}}
    \caption{Volumetric imaging of xenograft mouse model cadaver using HERCULES 45°. A tumor is present on the right hind leg of the mouse. a) shows a point 
    cloud volume rendering of the back of the mouse, whereas c) shows a cross-section of the volume.  The visible slice is also shown in d).}
    \label{fig:mouse}
\end{figure*}

\section{Results}

We simulated the PSF at a depth of 50 mm for a 128 $\times$ 128 array. In Figure \ref{fig:simulatedPsf} we see that all the imaging methods create well-resolved
points in the elevational-axial plane. RCAs are susceptible to the effects of edge waves, which results in echoes below the main lobe (a-h)) 
\cite{Jensen:2022:AnatomicRCA}. The axial and elevational resolutions of the various imaging methods are nearly identical, as they are primarily dependent on 
the wavelength of the ultrasound waves, and the physical properties of the array. On the other hand, the there is some variation laterally; While HERCULES sees 
no difference compared to it's elevational properties (i-j)), the use of tilted plane waves or virtual line sources creates significant differences (k-l)). We 
also measured the ratio of energy within 2.5 wavelengths of the point location, to that outside, and found that HERCULES maintained a higher inner energy ratio. 
This suggests that HERCULES will be less susceptible to sidelobes and be able to attain higher contrast when imaging complex structures. These results are 
summarized in Table \ref{table:simResults}.

In Figure \ref{fig:resolution} we validate our resolution measurements by imaging a series of line targets. Here, we obtain similar results from our 
simulations. We can see well resolved points in every imaging scheme in the xz plane and lines in the yz plane.  In the yz plane we are unable to image beyond 
the shadow of the aperture, except for in the case of HERCULES 45° (b)). We additionally show how lateral resolution varies with depth in Figure 
\ref{fig:resolutionMetric} by transmitting on the columns.  As expected, the resolution degrades as we go deeper as the relative aperture decreases. TPW is the 
most susceptible to this effect in its transmit plane. We characterize the imaging methods' elevational resolution by transmitting on the rows, and we can see
that there is effectively no difference between methods.

While HERCULES' effective imaging function benefits from not needing to assume the separability of its transmit and receive apertures, its primary advantage is 
the ability to image cylindrical fields of view. In Figure \ref{fig:cyst} we image anechoic cysts and assess the contrast of our imaging schemes, 
the results of which are shown in Figure \ref{fig:cystMetrics}.  We can see that all the imaging methods have comparable image quality and penetration, 
with a slight decrease in HERCULES 45°, but this is to be expected, as the transmitted energy is dispersed over a larger area. Secondly, the visibility of cysts, 
beyond the shadow of the aperture dramatically decreases, again with the exception of HERCULES 45° (Figure \ref{fig:cyst} b)). This is further corroborated
quantitatively by Figure \ref{fig:cystMetrics} d) where we obtain higher gCNR values for HERCULES 45° despite it being the lowest in a-c). The supplementary 
material contains flythroughs of these comparisons. It also contains a sector scan that shows how we can continuously vary the viewing region by changing 
the transmit delay profile. If HERCULES is used with a diverging acoustic lens, it would even be possible to image a full pyramidal volume, something that would 
normally only be achievable with a matrix, array; but even a matrix array cannot do this at high resolution due to channel count requirements.

\definecolor{headerColor}{rgb}{0.8,0.8,0.8}
\definecolor{rowColor1}{rgb}{1.0,1.0,1.0}
\definecolor{rowColor2}{rgb}{0.95,0.95,0.95}
\rowcolors{1}{rowColor1}{rowColor2}

\newcolumntype{H}[1]{>{\hsize=#1\hsize}X}

\begin{table*}
    \centering
    \caption{Imaging Scheme Comparison}
    \label{table:methodComparison}
    \begin{tabularx}{1.0\linewidth}{|H{0.8}|H{1.0}|H{1.0}|H{1.1}|H{1.1}|}

        \hline \rowcolor{headerColor}
        \textbf{Acquisition}                &
        \textbf{VLS}                        &
        \textbf{TPW}                        &
        \textbf{FORCES\slash uFORCES}       &
        \textbf{HERCULES}                   \\
        \hline
        References	                                                                &
        \cite{Rasmussen:2015:3DRCA1}	                                            &
        \cite{Chen:2018:RCA3DPW}	                                                &
        \cite{Ceroici:2017:Forces,Ceroici:2019:ForcesExp,Sobhani:2022:uForces}      &
        N/A                                                                         \\
        \hline
        Possible with Row-Column Arrays &
        Yes	            &
        Yes	            &
        No	            &
        No              \\
        \hline
        \raggedright Possible with \linebreak  Electrostrictive \linebreak TOBE Arrays  &
        Yes	            &
        Yes	            &
        Yes	            &
        Yes             \\
        \hline
        Description	&

        -- Virtual line sources emissions on columns while receiving on all rows (or vice versa).

        -- Synthetic transmit focusing in azimuth, dynamic receive focusing in elevation (or vice verse).	&

        -- Tilted plane wave emissions with reception on rows then columns.	&

        -- Transmit Aperture encoded multi-static synthetic aperture imaging with electronically-steerable scan plane and elevation focus. &

        -- Plane or cylindrically diverging wave emissions, reception on every element using Hadamard aperture-encoded readout (HERO). \\

        \hline

        Implementation  &	
        -- Tx on columns with walking cylindrical wave emissions, receive on rows (or vice versa). &

        -- Tx on columns to form tilted plane wave emissions, receive on all rows, then reverse roles of rows and columns. &	

        -- Tx on rows while biasing and Rx on columns (or vice versa). Invert polarity of Rx signals when negatively biased, aperture-decode and beamform. &	

        -- Bias and Tx on columns, inverting Tx polarity when negatively biased, Rx from rows. Aperture decode to emulate fully-wired 2D array on reception. Can reverse roles of rows and columns. \\

        \hline

        Capabilities &

            -- 3D imaging

            -- Seperable row- and column focusing  &

            -- 3D imaging

            -- Seperable row- and column focusing  &

            --  Very high-quality B-scan beyond the shadow of the aperture

            -- 2D images with transmit and receive focusing (2-way focusing) everywhere in-plane

            -- Electronically-steerable scan plane and elevation focus  &

            -- 3D imaging beyond the shadow of the aperture

            -- Similar image quality to VLS and TPW methods but larger FOV

            -- Spherical receive focusing everywhere in the volume \\

        \hline

        Limitations	&

            -- One-way elevation focusing

            -- One-way azimuthal focusing

            -- Imaging only beneath the shadow of the aperture &

            -- One-way elevation focusing

            -- One-way azimuthal focusing

            -- Imaging only beneath the shadow of the aperture &

            -- One-way elevation focusing

            -- Transmit focusing limited to signal elevational focal depth &

            --  Full pyramidal volumes not yet achieved

            -- Unfocused transmissions lead to one-way focusing (receive focusing only)  \\

        \hline

        Hardware Needed &	
        -- Research Ultrasound Platform + RCA	&
        -- Research Ultrasound Platform + RCA	&
        -- Research Ultrasound Platform + TOBE Array + Biasing Electronics	&
        -- Research Ultrasound Platform + TOBE Array + Biasing Electronics \\

        \hline
    \end{tabularx}
\end{table*}

We also show that HERCULES is able to effectively image real tissue. In Figure \ref{fig:mouse} we image a xenograft mouse model cadaver, with all imaging 
methods. A tumor can be seen on the hind leg of the mouse.  In all these volumetric imaging methods sidelobes from the strong scattering against the surface of 
the mouse and its bones create artifacts that reduce the overall quality of the volume. These artifacts are characteristic of the volumetric imaging limitations
of row-column arrays. We show a slice from the HERCULES volume (a)) as proper 3D visualization of ultrasound data is a non-trivial problem. It is difficult to 
make 2D arrays suitable for small animal volumetric imaging, but TOBE arrays have the potential to fill this space.

For ultrafast imaging, limiting transmit counts is important to minimize the effect of motion artifacts and allow us to perform motion compensation, as well as 
enable applications like flow imaging which require the collection of many volumes per second \cite{Denarie:2013:CoherentPWRapid, 
Pustovalov:2022:MotionCompensationBloodFlow}. All the previous images were made with 128 transmit events. With VLS \& TPW we can simply reduce the amount of 
transmits, use sparser effective transmit apertures and create lower quality volumes. This is less trivial for HERCULES, as Hadamard matrices do not exist for 
an arbitrary number of elements, we will no longer be decoding the entire array. However, that is not to say that there are no other options to reduce transmit 
count; grouping elements into bins and using a smaller Hadamard matrix is one possible solution. Active motion compensation can also potentially be used to 
alleviate motion artifacts.

Figure \ref{fig:phantomVolume} demonstrates a unique advantage of TOBE arrays; large volumes of tissues can be scanned using HERCULES, and then high quality and 
resolution images can be obtained using a companion imaging method Fast-Orthogonal-Row-Column-Electronic-Scanning (FORCES) \cite{Ceroici:2017:Forces, 
Ceroici:2019:ForcesExp}. Table \ref{table:methodComparison} summarizes the differences between TOBE imaging methods and regular RCA methods.

\section{Conclusion}

In this work we were able to demonstrate a new imaging method that leveraged the properties of bias-sensitive row column arrays with bias encoding to perform 
high resolution volumetric structural imaging at real-time rates without mechanical scanning. This method has comparable imaging quality to other row-column 
imaging techniques, with the additional unique benefit of being able to image beyond the shadow of the aperture.  These properties help to open the way for 
various clinical applications such as full organ imaging, and imaging volumes through limited apertures, such as cardiovascular imaging through the ribs, and 
intracranial imaging. Future work could include implementation of methods to reduce the required transmits to generate a volume. Volumetric flow, 
super-resolution, and elastography imaging could also build on this work.

\bibliographystyle{IEEEtran}
\bibliography{hercules}

\begin{thebibliography}{10}
\providecommand{\url}[1]{#1}
\csname url@samestyle\endcsname
\providecommand{\newblock}{\relax}
\providecommand{\bibinfo}[2]{#2}
\providecommand{\BIBentrySTDinterwordspacing}{\spaceskip=0pt\relax}
\providecommand{\BIBentryALTinterwordstretchfactor}{4}
\providecommand{\BIBentryALTinterwordspacing}{\spaceskip=\fontdimen2\font plus
\BIBentryALTinterwordstretchfactor\fontdimen3\font minus \fontdimen4\font\relax}
\providecommand{\BIBforeignlanguage}[2]{{%
\expandafter\ifx\csname l@#1\endcsname\relax
\typeout{** WARNING: IEEEtran.bst: No hyphenation pattern has been}%
\typeout{** loaded for the language `#1'. Using the pattern for}%
\typeout{** the default language instead.}%
\else
\language=\csname l@#1\endcsname
\fi
#2}}
\providecommand{\BIBdecl}{\relax}
\BIBdecl

\bibitem{Smith:1991:HighSpeedSystemA}
S.~Smith, H.~Pavy, and O.~von Ramm, ``High-speed ultrasound volumetric imaging system. i. transducer design and beam steering,'' \emph{IEEE Transactions on Ultrasonics, Ferroelectrics, and Frequency Control}, vol.~38, no.~2, pp. 100--108, 1991.

\bibitem{Welch:2001:Freehand3D}
\BIBentryALTinterwordspacing
J.~N. Welch, J.~A. Johnson, M.~R. Bax, R.~B. M.D., S.~S. M.D., T.~Krummel, and R.~Shahidi, ``{Real-time freehand 3D ultrasound system for clinical applications},'' in \emph{Medical Imaging 2001: Visualization, Display, and Image-Guided Procedures}, S.~K. Mun, Ed., vol. 4319, International Society for Optics and Photonics.\hskip 1em plus 0.5em minus 0.4em\relax SPIE, 2001, pp. 724 -- 730. [Online]. Available: \url{https://doi.org/10.1117/12.428120}
\BIBentrySTDinterwordspacing

\bibitem{Austeng:2002:Sparse2DArray3DImaging}
A.~Austeng and S.~Holm, ``Sparse 2-d arrays for 3-d phased array imaging - design methods,'' \emph{IEEE Transactions on Ultrasonics, Ferroelectrics, and Frequency Control}, vol.~49, no.~8, pp. 1073--1086, 2002.

\bibitem{Yen:2004:3DRectilinearMultiplex}
J.~Yen and S.~Smith, ``Real-time rectilinear 3-d ultrasound using receive mode multiplexing,'' \emph{IEEE Transactions on Ultrasonics, Ferroelectrics, and Frequency Control}, vol.~51, no.~2, pp. 216--226, 2004.

\bibitem{Chang:2015:GraphSegmentation3D}
\BIBentryALTinterwordspacing
H.~Chang, Z.~Chen, Q.~Huang, J.~Shi, and X.~Li, ``Graph-based learning for segmentation of 3d ultrasound images,'' \emph{Neurocomputing}, vol. 151, pp. 632--644, 2015. [Online]. Available: \url{https://www.sciencedirect.com/science/article/pii/S0925231214013873}
\BIBentrySTDinterwordspacing

\bibitem{Downey:1995:3DVascularDoppler}
\BIBentryALTinterwordspacing
D.~B. Downey and A.~Fenster, ``Vascular imaging with a three-dimensional power doppler system.'' \emph{American Journal of Roentgenology}, vol. 165, no.~3, pp. 665--668, 1995, pMID: 7645492. [Online]. Available: \url{https://doi.org/10.2214/ajr.165.3.7645492}
\BIBentrySTDinterwordspacing

\bibitem{Huang:2015:3DLinearStrain}
Q.~Huang, B.~Xie, P.~Ye, and Z.~Chen, ``Correspondence - 3-d ultrasonic strain imaging based on a linear scanning system,'' \emph{IEEE Transactions on Ultrasonics, Ferroelectrics, and Frequency Control}, vol.~62, no.~2, pp. 392--400, 2015.

\bibitem{Lockwood:1998:3DSparseSA}
G.~Lockwood, J.~Talman, and S.~Brunke, ``Real-time 3-d ultrasound imaging using sparse synthetic aperture beamforming,'' \emph{IEEE Transactions on Ultrasonics, Ferroelectrics, and Frequency Control}, vol.~45, no.~4, pp. 980--988, 1998.

\bibitem{vonRamm:1991:HighSpeedSystemB}
O.~von Ramm, S.~Smith, and H.~Pavy, ``High-speed ultrasound volumetric imaging system. ii. parallel processing and image display,'' \emph{IEEE Transactions on Ultrasonics, Ferroelectrics, and Frequency Control}, vol.~38, no.~2, pp. 109--115, 1991.

\bibitem{Seo:2009:RCA}
C.~H. Seo and J.~T. Yen, ``A 256 x 256 2-d array transducer with row-column addressing for 3-d rectilinear imaging,'' \emph{IEEE Transactions on Ultrasonics, Ferroelectrics, and Frequency Control}, vol.~56, no.~4, pp. 837--847, 2009.

\bibitem{Rasmussen:2013:3DRCASimulation}
\BIBentryALTinterwordspacing
M.~F. Rasmussen and J.~A. Jensen, ``3d ultrasound imaging performance of a row-column addressed 2d array transducer: a simulation study,'' in \emph{Medical Imaging 2013: Ultrasonic Imaging, Tomography, and Therapy}, J.~G. Bosch and M.~M. Doyley, Eds., vol. 8675, International Society for Optics and Photonics.\hskip 1em plus 0.5em minus 0.4em\relax SPIE, 2013, p. 86750C. [Online]. Available: \url{https://doi.org/10.1117/12.2007083}
\BIBentrySTDinterwordspacing

\bibitem{Wong:2014:MicromachinedRCA}
\BIBentryALTinterwordspacing
L.~L. Wong, A.~I. Chen, Z.~Li, A.~S. Logan, and J.~T. Yeow, ``A row-column addressed micromachined ultrasonic transducer array for surface scanning applications,'' \emph{Ultrasonics}, vol.~54, no.~8, pp. 2072--2080, 2014. [Online]. Available: \url{https://www.sciencedirect.com/science/article/pii/S0041624X14001929}
\BIBentrySTDinterwordspacing

\bibitem{Awad:2009:3DSpatialCompoundRCA}
\BIBentryALTinterwordspacing
S.~I. Awad and J.~T. Yen, ``3-d spatial compounding using a row-column array,'' \emph{Ultrasonic Imaging}, vol.~31, no.~1, pp. 120--130, 2009, pMID: 19630253. [Online]. Available: \url{https://doi.org/10.1177/016173460903100103}
\BIBentrySTDinterwordspacing

\bibitem{Flesch:2017:4DUltrafastRCA}
\BIBentryALTinterwordspacing
M.~Flesch, M.~Pernot, J.~Provost, G.~Ferin, A.~Nguyen-Dinh, M.~Tanter, and T.~Deffieux, ``4d in vivo ultrafast ultrasound imaging using a row-column addressed matrix and coherently-compounded orthogonal plane waves,'' \emph{Physics in Medicine \& Biology}, vol.~62, no.~11, p. 4571, may 2017. [Online]. Available: \url{https://dx.doi.org/10.1088/1361-6560/aa63d9}
\BIBentrySTDinterwordspacing

\bibitem{Jørgensen:2023:RCABeamformer}
L.~T. Jørgensen, S.~K. Præsius, M.~B. Stuart, and J.~A. Jensen, ``Row-column beamformer for fast volumetric imaging,'' \emph{IEEE Transactions on Ultrasonics, Ferroelectrics, and Frequency Control}, vol.~70, no.~7, pp. 668--680, 2023.

\bibitem{Shattuck:1984:Explososcan}
\BIBentryALTinterwordspacing
D.~P. Shattuck, M.~D. Weinshenker, S.~W. Smith, and O.~T. von Ramm, ``Explososcan: A parallel processing technique for high speed ultrasound imaging with linear phased arrays,'' \emph{The Journal of the Acoustical Society of America}, vol.~75, no.~4, pp. 1273--1282, 04 1984. [Online]. Available: \url{https://doi.org/10.1121/1.390734}
\BIBentrySTDinterwordspacing

\bibitem{Jain-Yu:1997:LimitedDiffractionBeams}
J.-Y. Lu, ``2d and 3d high frame rate imaging with limited diffraction beams,'' \emph{IEEE Transactions on Ultrasonics, Ferroelectrics, and Frequency Control}, vol.~44, no.~4, pp. 839--856, 1997.

\bibitem{Jain-Yu:1998:LimitedDiffractionBeamsExperiments}
------, ``Experimental study of high frame rate imaging with limited diffraction beams,'' \emph{IEEE Transactions on Ultrasonics, Ferroelectrics, and Frequency Control}, vol.~45, no.~1, pp. 84--97, 1998.

\bibitem{Sauvage:2018:4DUltrafastDopplerRCA}
\BIBentryALTinterwordspacing
J.~Sauvage, M.~Flesch, G.~Férin, A.~Nguyen-Dinh, J.~Porée, M.~Tanter, M.~Pernot, and T.~Deffieux, ``A large aperture row column addressed probe for in vivo 4d ultrafast doppler ultrasound imaging,'' \emph{Physics in Medicine \& Biology}, vol.~63, no.~21, p. 215012, oct 2018. [Online]. Available: \url{https://dx.doi.org/10.1088/1361-6560/aae427}
\BIBentrySTDinterwordspacing

\bibitem{Jensen:2022:RCALinearComparison}
J.~A. Jensen, M.~Schou, M.~L. Ommen, K.~Steenberg, E.~V. Thomsen, B.~G. Tomov, N.~S. Panduro, C.~M. Sorensen, and M.~B. Stuart, ``Synthetic aperture high quality b-mode imaging with a row-column array compared to linear array imaging,'' in \emph{2022 IEEE International Ultrasonics Symposium (IUS)}, 2022, pp. 1--4.

\bibitem{Jensen:2022:AnatomicRCA}
J.~A. Jensen, M.~Schou, L.~T. Jørgensen, B.~G. Tomov, M.~B. Stuart, M.~S. Traberg, I.~Taghavi, S.~H. Øygaard, M.~L. Ommen, K.~Steenberg, E.~V. Thomsen, N.~S. Panduro, M.~B. Nielsen, and C.~M. Sørensen, ``Anatomic and functional imaging using row-column arrays,'' \emph{IEEE Transactions on Ultrasonics, Ferroelectrics, and Frequency Control}, vol.~69, no.~10, pp. 2722--2738, 2022.

\bibitem{Wu:2024:3DTranscranialDynamicUltrasound}
\BIBentryALTinterwordspacing
A.~Wu, J.~Porée, G.~Ramos-Palacios, C.~Bourquin, N.~Ghigo, A.~Leconte, P.~Xing, A.~F. Sadikot, M.~Chassé, and J.~Provost, ``3d transcranial dynamic ultrasound localization microscopy in the mouse brain using a row-column array,'' 2024. [Online]. Available: \url{https://arxiv.org/abs/2406.01746}
\BIBentrySTDinterwordspacing

\bibitem{Sherrit:1998:Electrostrictive}
\BIBentryALTinterwordspacing
S.~Sherrit and B.~K. Mukherjee, ``{Electrostrictive materials: characterization and applications for ultrasound},'' in \emph{Medical Imaging 1998: Ultrasonic Transducer Engineering}, K.~K. Shung, Ed., vol. 3341, International Society for Optics and Photonics.\hskip 1em plus 0.5em minus 0.4em\relax SPIE, 1998, pp. 196 -- 207. [Online]. Available: \url{https://doi.org/10.1117/12.308000}
\BIBentrySTDinterwordspacing

\bibitem{Ceroici:2017:Forces}
C.~Ceroici, T.~Harrison, and R.~J. Zemp, ``Fast orthogonal row-column electronic scanning with top-orthogonal-to-bottom electrode arrays,'' \emph{IEEE Transactions on Ultrasonics, Ferroelectrics, and Frequency Control}, vol.~64, no.~6, pp. 1009--1014, 2017.

\bibitem{Ceroici:2019:ForcesExp}
C.~Ceroici, K.~Latham, B.~A. Greenlay, J.~A. Brown, and R.~J. Zemp, ``Fast orthogonal row-column electronic scanning experiments and comparisons,'' \emph{IEEE Transactions on Ultrasonics, Ferroelectrics, and Frequency Control}, vol.~66, no.~6, pp. 1093--1101, 2019.

\bibitem{Sobhani:2022:uForces}
M.~R. Sobhani, M.~Ghavami, A.~K. Ilkhechi, J.~Brown, and R.~Zemp, ``Ultrafast orthogonal row-column electronic scanning (uforces) with bias-switchable top-orthogonal-to-bottom electrode 2-d arrays,'' \emph{IEEE Transactions on Ultrasonics, Ferroelectrics, and Frequency Control}, vol.~69, no.~10, pp. 2823--2836, 2022.

\bibitem{Ceroici:2018:Photoacoustic}
C.~Ceroici, K.~Latham, R.~Chee, B.~Greenlay, Q.~Barber, J.~A. Brown, and R.~Zemp, ``3d photoacoustic imaging using hadamard-bias encoding with a crossed electrode relaxor array,'' \emph{Opt. Lett.}, vol.~43, no.~14, pp. 3425--3428, Jul 2018.

\bibitem{Ceroici:Thesis}
C.~Ceroici, ``Novel 3d ultrasound imaging techniques using top-orthogonal- to-bottom-electrode (tobe) arrays,'' Ph.D. dissertation, University of Alberta, 2018.

\bibitem{Rasmussen:2015:3DRCA1}
M.~F. Rasmussen, T.~L. Christiansen, E.~V. Thomsen, and J.~A. Jensen, ``3-d imaging using row-column-addressed arrays with integrated apodization - part i: apodization design and line element beamforming,'' \emph{IEEE Transactions on Ultrasonics, Ferroelectrics, and Frequency Control}, vol.~62, no.~5, pp. 947--958, 2015.

\bibitem{Chen:2018:RCA3DPW}
K.~Chen, B.~C. Lee, K.~E. Thomenius, B.~T. Khuri-Yakub, H.-S. Lee, and C.~G. Sodini, ``A column-row-parallel ultrasound imaging architecture for 3-d plane-wave imaging and tx second-order harmonic distortion reduction,'' \emph{IEEE Transactions on Ultrasonics, Ferroelectrics, and Frequency Control}, vol.~65, no.~5, pp. 828--843, 2018.

\bibitem{FieldII:Theory}
J.~Jensen and N.~Svendsen, ``Calculation of pressure fields from arbitrarily shaped, apodized, and excited ultrasound transducers,'' \emph{IEEE Transactions on Ultrasonics, Ferroelectrics, and Frequency Control}, vol.~39, no.~2, pp. 262--267, 1992.

\bibitem{FieldII:Program}
J.~Jensen, ``\BIBforeignlanguage{English}{Field: A program for simulating ultrasound systems},'' \emph{\BIBforeignlanguage{English}{Medical \& Biological Engineering \& Computing}}, vol.~34, no. sup. 1, pp. 351--353, 1997, 10th Nordic-Baltic Conference on Biomedical Imaging ; Conference date: 09-06-1996 Through 13-06-1996.

\bibitem{Sampaleanu:2014:CMUT_TOBE_Ultrasound}
A.~Sampaleanu, P.~Zhang, A.~Kshirsagar, W.~Moussa, and R.~J. Zemp, ``Top-orthogonal-to-bottom-electrode (tobe) cmut arrays for 3-d ultrasound imaging,'' \emph{IEEE Transactions on Ultrasonics, Ferroelectrics, and Frequency Control}, vol.~61, no.~2, pp. 266--276, 2014.

\bibitem{Chee:2014:CMUT_TOBE_Photoacoustic}
R.~K. Chee, A.~Sampaleanu, D.~Rishi, and R.~J. Zemp, ``Top orthogonal to bottom electrode (tobe) 2-d cmut arrays for 3-d photoacoustic imaging,'' \emph{IEEE Transactions on Ultrasonics, Ferroelectrics, and Frequency Control}, vol.~61, no.~8, pp. 1393--1395, 2014.

\bibitem{Ilkhechi:2023:BiasSwitching}
A.~K. Ilkhechi, R.~Palamar, M.~R. Sobhani, D.~Dahunsi, C.~Ceroici, M.~Ghavami, J.~Brown, and R.~Zemp, ``High-voltage bias-switching electronics for volumetric imaging using electrostrictive row–column arrays,'' \emph{IEEE Transactions on Ultrasonics, Ferroelectrics, and Frequency Control}, vol.~70, no.~4, pp. 324--335, 2023.

\bibitem{Jensen:2006:SyntheticAperture}
\BIBentryALTinterwordspacing
J.~A. Jensen, S.~I. Nikolov, K.~L. Gammelmark, and M.~H. Pedersen, ``Synthetic aperture ultrasound imaging,'' \emph{Ultrasonics}, vol.~44, pp. e5--e15, 2006, proceedings of Ultrasonics International (UI'05) and World Congress on Ultrasonics (WCU). [Online]. Available: \url{https://www.sciencedirect.com/science/article/pii/S0041624X06003374}
\BIBentrySTDinterwordspacing

\bibitem{Nikolov:2006:SyntheticApertureChallenges}
S.~I. Nikolov, B.~G. Tomov, and J.~A. Jensen, ``Real-time synthetic aperture imaging: opportunities and challenges,'' in \emph{2006 Fortieth Asilomar Conference on Signals, Systems and Computers}, 2006, pp. 1548--1552.

\bibitem{Misaridis:2005:CodedExcitationA}
T.~Misaridis and J.~Jensen, ``Use of modulated excitation signals in medical ultrasound. part i: basic concepts and expected benefits,'' \emph{IEEE Transactions on Ultrasonics, Ferroelectrics, and Frequency Control}, vol.~52, no.~2, pp. 177--191, 2005.

\bibitem{Misaridis:2005:CodedExcitationB}
------, ``Use of modulated excitation signals in medical ultrasound. part ii: design and performance for medical imaging applications,'' \emph{IEEE Transactions on Ultrasonics, Ferroelectrics, and Frequency Control}, vol.~52, no.~2, pp. 192--207, 2005.

\bibitem{Misaridis:2005:CodedExcitationC}
------, ``Use of modulated excitation signals in medical ultrasound. part iii: high frame rate imaging,'' \emph{IEEE Transactions on Ultrasonics, Ferroelectrics, and Frequency Control}, vol.~52, no.~2, pp. 208--219, 2005.

\bibitem{Fink:1992:1:TimeReversalA}
M.~Fink, ``Time reversal of ultrasonic fields. i. basic principles,'' \emph{IEEE Transactions on Ultrasonics, Ferroelectrics, and Frequency Control}, vol.~39, no.~5, pp. 555--566, 1992.

\bibitem{MATLAB}
\BIBentryALTinterwordspacing
T.~M. Inc., ``Matlab version: 9.13.0 (r2022b),'' Natick, Massachusetts, United States, 2022. [Online]. Available: \url{https://www.mathworks.com}
\BIBentrySTDinterwordspacing

\bibitem{Rodriguez-Molares:2020:gCNR}
A.~Rodriguez-Molares, O.~M.~H. Rindal, J.~D'hooge, S.-E. Måsøy, A.~Austeng, M.~A. Lediju~Bell, and H.~Torp, ``The generalized contrast-to-noise ratio: A formal definition for lesion detectability,'' \emph{IEEE Transactions on Ultrasonics, Ferroelectrics, and Frequency Control}, vol.~67, no.~4, pp. 745--759, 2020.

\bibitem{Denarie:2013:CoherentPWRapid}
B.~Denarie, T.~A. Tangen, I.~K. Ekroll, N.~Rolim, H.~Torp, T.~Bjåstad, and L.~Lovstakken, ``Coherent plane wave compounding for very high frame rate ultrasonography of rapidly moving targets,'' \emph{IEEE Transactions on Medical Imaging}, vol.~32, no.~7, pp. 1265--1276, 2013.

\bibitem{Pustovalov:2022:MotionCompensationBloodFlow}
\BIBentryALTinterwordspacing
V.~Pustovalov, D.~H. Pham, J.~P. Remenieras, and D.~Kouam{\'e}, ``{Motion compensation for the estimation of high-resolution blood flow in ultrafast ultrasound imaging},'' in \emph{Medical Imaging 2022: Ultrasonic Imaging and Tomography}, N.~Bottenus and N.~V. Ruiter, Eds., vol. 12038, International Society for Optics and Photonics.\hskip 1em plus 0.5em minus 0.4em\relax SPIE, 2022, p. 120380D. [Online]. Available: \url{https://doi.org/10.1117/12.2607057}
\BIBentrySTDinterwordspacing

\end{thebibliography}

\end{document}